\documentclass[10pt]{iopart}
\usepackage{graphics,graphicx,epsfig,color,hyperref,latexsym,float,amssymb}

\begin{document}

\title{Magnetic Order and Mott Transition on the Checkerboard Lattice}

\author{Nyayabanta Swain and Pinaki Majumdar}

\address{Harish-Chandra  Research Institute,
Chhatnag Road, Jhunsi, Allahabad 211019, India \\
Homi Bhabha National Institute, Training School Complex, 
Anushakti Nagar, Mumbai 400085, India}

\ead{nyaya@hri.res.in, pinaki@hri.res.in}


\begin{abstract}
The checkerboard lattice, with alternating `crossed' plaquettes, serves as the 
two dimensional analog of the pyrochlore lattice. The corner sharing plaquette 
structure leads to a hugely degenerate ground state, and no magnetic order, for 
classical spins with short range antiferromagnetic interaction. For the half-filled 
Hubbard model on this structure, however, we find that the Mott insulating phase 
involves virtual electronic processes that generate longer range and 
multispin couplings. These couplings lift the degeneracy, 
selecting a `flux like' state in the Mott insulator. 
Increasing temperature leads, strangely, to a sharp crossover from this 
state to a `120 degree' correlated state and then a paramagnet. Decrease in the 
Hubbard repulsion drives the system towards an insulator-metal transition - the 
moments reduce, and a spin disordered state wins over the flux state. Near the
insulator-metal transition the electron system displays a pseudogap extending 
over a large temperature window.
\end{abstract}

\maketitle

\section{INTRODUCTION}

\begin{figure}[b]
\centering{
\includegraphics[width=4.2cm,height=4.2cm,angle=0]{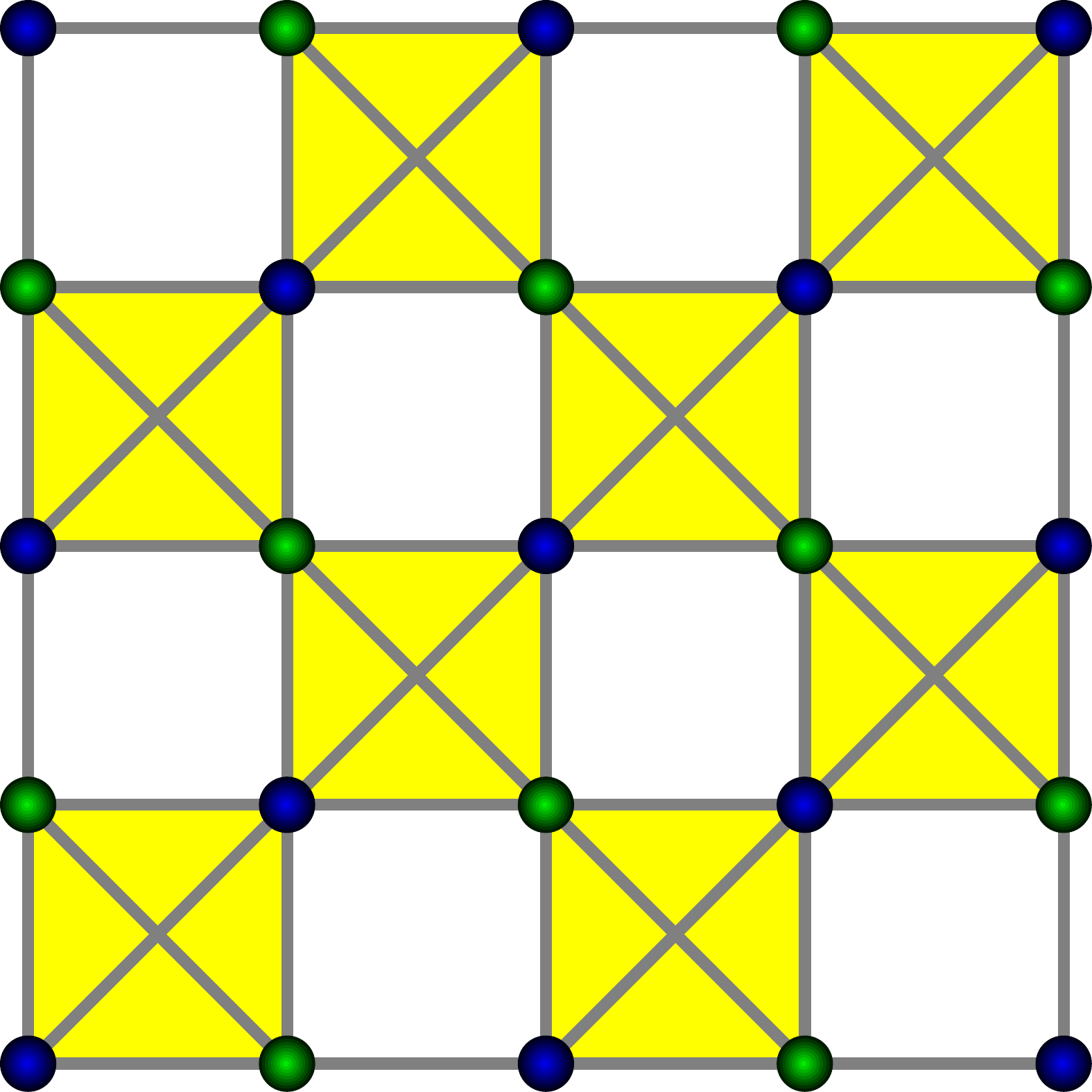}
}
\caption{\label{chkb_lattice}
Color online:
The checkerboard lattice with alternating `empty' and crossed
plaquettes. All hopping amplitudes (diagonal and axial) are equal.
The structural unit cell contains  two atoms.
Each crossed plaquette, in isolation, can be viewed as a tetrahedron.
}
\end{figure}

Frustrated magnets arise due to the coupling between 
electrons localised on non bipartite lattices
\cite{fr-latt1,fr-latt2,fr-latt3}.
The localisation stems from electron correlation,
a concrete example being the Mott insulating phase 
of the half filled Hubbard model \cite{MIT_Mott}.
The $U/t \gg 1$ limit in such cases, where
$U$ is the Hubbard repulsion and $t$ the typical kinetic
scale, involves  virtual hopping 
of the electrons, and induces  
antiferromagnetic exchange.
With weakening $U/t$ 
the electrons `delocalise' over progressively 
longer distance and mediate longer range couplings
\cite{MSE1,MSE2}.  These additional 
couplings are crucial in deciding the physics 
when the $U/t \gg 1$ Heisenberg 
limit involves 
a macroscopically degenerate ground
state. The checkerboard lattice, Fig.\ref{chkb_lattice}, 
is in this category \cite{chkb_cHAF}.

Two  complications arise with decreasing
$U/t$: (a)~the size of the moments diminish as the
system heads towards the Mott transition, 
and (b)~the range
of electron hops increase and the exchange processes get
harder to quantify. 
Near the insulator-metal transition (IMT)
 the magnetic correlations on the frustrated 
lattice, and their impact on electronic properties,
have to be worked out self-consistently.
This has indeed been attempted for various
frustrated lattices, 
{\it e.g}, the edge-shared triangular 
\cite{trlatt0,trlatt1,trlatt2,trlatt3,trlatt4,rt-trlatt} 
and FCC \cite{fcc1,rt-fcc,fcc3} lattices,
the corner shared kagome \cite{kagome1,kagome2}
and pyrochlore \cite{pyr1,pyr2}
lattices. Surprisingly, very little
attention has been given to the  
frustrated checkerboard latice.

For the checkerboard lattice the Heisenberg antiferromagnet
is well understood
\cite{chkb_cHAF,chkb_qHAF1,chkb_qHAF2,chkb_qHAF3,chkb_qHAF4,
chkb_qHAF5,chkb_qHAF6,chkb_qHAF7,chkb_qHAF8,chkb_qHAF9,chkb_qHAF10}.
The classical ground state is macroscopically
degenerate \cite{chkb_cHAF} and there is no order at
any temperature.
The 
quantum, $S=1/2$, case is argued to be a
plaquette valence-bond crystal 
\cite{chkb_qHAF1,chkb_qHAF2,chkb_qHAF3,chkb_qHAF4,
chkb_qHAF5,chkb_qHAF6,chkb_qHAF7,chkb_qHAF8,chkb_qHAF9,chkb_qHAF10}
- the product of singlets
on the uncrossed plaquettes.

There is limited work on the checkerboard Hubbard model,
focused mainly on the ground state.
One study \cite{mit_fujimoto} 
suggested that increasing $U/t$ 
leads to a
transition from  the  semi-metallic band state to
a charge-ordered insulator, and then to a 
magnetically disordered Mott insulator, while another 
reports a first order transition 
from the semi-metal 
to an insulating state with plaquette magnetic 
order \cite{mit_yoshioka}.
 
The varying results, based on different methods 
\cite{mit_fujimoto,mit_yoshioka,chkb_yokoyama}, 
leave some questions unanswered: 
(i)~What magnetic ground state emerges within a Hartree-Fock (HF)
scheme, non trivial here since the HF state may be disordered.
(ii)~What are the effects of thermal fluctuation on this magnetic
state?
and (iii)~What is the impact of the magnetic order, and
fluctuations, on the electronic properties near the IMT?
We address these questions using a combination of 
Hartree-Fock theory for the
ground state and an auxiliary field based 
Monte Carlo to handle thermal fluctuations.
Our results reveal a wide variety of magnetic and 
spectral regimes on this lattice, summarised below.

(1).~{\it Strong coupling:} Deep in the Mott phase
the Hubbard model selects out a flux like 
state from the infinitely degenerate manifold 
of the checkerboard Heisenberg model. 
This persists as the low temperature state down to $U \sim 5.3t$. 
Increasing temperature promotes a $120$ degree 
spin arrangement, before the final loss of order.
(2).~{\it Weak coupling:} Small {\it disordered}
moments persist below the 
insulator-metal transition (at $U \sim 5.3t$) down 
to $U \sim 3t$. 
(3).~{\it Electronic state:} 
The electrons are gapped in the flux phase, 
but the crossover to the $120$ degree state leads
to a reduction of the gap. 
As $U/t$ reduces, the $120$ degree phase becomes 
pseudogapped.

\section{MODEL AND METHOD}

We  consider the single band 
Hubbard model on the checkerboard lattice,
$$
H = 
\sum_{ij, \sigma} ( t_{ij} - \mu \delta_{ij} ) 
c_{i\sigma}^{\dagger}c_{j\sigma}
+ U \sum_{i} n_{i\uparrow} n_{i\downarrow}
$$
$t_{ij} = -t$ for the nearest neighbour (NN) 
axial as well as diagonal bonds.
$U > 0$ is the interaction strength and $\mu$ is chosen
to keep the system at electron density $n=1$.
We set $t=1$ and measure all energies in this unit.

We follow an approach originally suggested by Hubbard 
\cite{HS1} to obtain a rotation invariant 
auxiliary field decomposition \cite{HS2,HS3}
of the interaction term. 
Retaining rotation invariance
{\it and} reproducing Hartree-Fock theory at 
saddle point requires the introduction of a
three dimansional auxiliary vector
field, ${\vec m}_i(\tau)$, and a scalar field $\phi_i(\tau)$.
We treat $\phi$ as independent of $i, \tau$ 
(as would happen at saddle point) while for
${\vec m}_i$ we  retain the 
spatial fluctuations but not the imaginary time dependence.
In this limit the partition function can be written in terms
of the effective Hamiltonian \cite{rt-trlatt}: 
$$
H_{eff} = 
\sum_{ij, \sigma} t_{ij} c_{i\sigma}^{\dagger}c_{j\sigma}
- {\tilde \mu} N
-{ U \over 2} \sum_i {\vec m}_i.{\sigma}_i
+
\frac{U}{4} \sum_{i} \vec{m_{i}}^{2}
$$
where $\tilde{\mu} = \mu - (U/2)$ and
${\vec \sigma}_i$ is the electron spin operator. 
The distribution function of the $\{ {\vec m}_i \}$ is 
obtained by tracing over the fermions:
$$
P\{ {\vec m}_i \} \sim
Tr_{c,c^{\dagger}} e^{- \beta H_{eff}\{ {\vec m}_i \}}
$$
with $\beta = 1/(k_B T)$, where $T$ is the temperature
and we set $k_B =1$.
The equations for $H_{eff}$ and $P\{ {\vec m}_i \}$ define
a self-consistent loop.

To gain some insight it is helpful
to separate three regimes. 
(1)~As $T \rightarrow 0$ the problem reduces to 
{\it minimising} the energy of $H_{eff}$ 
with respect to the ${\vec m}_i$.
The minimisation is equivalent to the 
HF condition ${\vec m}_i = \langle {\vec \sigma}_i \rangle$.
So at $T=0$ this approach just reproduces mean field theory (without
any assumption, however, about the spatial organisation of the
${\vec m}_i$).  (2)~At low finite $T$, 
the crucial low energy angular fluctuations
of ${\vec m}_i$ come into play, revealing the thermal correlation
scale of the moments.
(3)~At high $T$, where the ${\vec m}_i$ are
essentially random, the auxiliary field 
mimics the effect of electron-electron
interaction mainly by preventing double occupancy in
the large $U/t$ regime.

We use two approaches to study $H_{eff}$. 
(i)~At finite $T$ we use a Monte Carlo (MC) approach, 
using a cluster algorithm \cite{tca_pinaki_sanjeev}
to generate equilibrium configurations of the ${\vec m}_i$. 
We typically use a $ N = 24\times 24$ 
lattice with $8\times 8$ update clusters. 
We anneal down from $T = 0.1t$ (since we see no
correlations above that temperature) and use $10^4$ MC sweeps
per temperature, going down to $T = 10^{-4}t$.
Physical properties are averaged over $\sim 100$
configurations at each $T$.
(ii)~At $T=0$ we use a variational scheme, 
trying out a family of periodic ${\vec m}_i$ configurations
(both planar and non-planar) 
and cross check our results with the Monte Carlo 
based annealing of the ${\vec m}_i$. 

To characterise the magnetic state we calculate
the following indicators with the equilibrium 
MC configurations.
\begin{eqnarray}
S(\vec{q}) &=&  \frac{1}{N^{2}} \sum_{ij}
\langle \vec{m_{i}}.\vec{m_{j}} \rangle
e^{i\vec{q} \cdot (\vec{R_{i}} - \vec{R_{j}})}
\cr
\tau_{avg} & = & \frac{1}{N} \sum_{i} \int_{0}^{t_{max}} dt'
\langle \vec{m_{i}}(0).\vec{m_{i}}(t') \rangle
\cr 
P(m) & = & \frac{1}{N} \sum_i \langle \delta(m - \vert {\vec m}_i \vert)
\rangle
\nonumber
\end{eqnarray}
In the expressions above $N$ is the system size and
the angular brackets stand for thermal average.
In sequence, (i)~$S(\vec{q})$  is the  
thermally averaged magnetic structure factor.
The onset of rapid growth in $S(\vec{q})$
at some wavevector $\vec{Q}$
indicates magnetic ordering.
In the thermodynamic limit,
there would be no `true' long-range magnetic ordering 
at finite temperature 
in two dimensions (2D). Our magnetic orders refer to 
growing magnetic correlation length scales beyond system sizes.
We have checked the size dependence of the various 
temperature scales associated with these crossovers 
within the MC. Our MC runs on $16 \times 16$, $24 \times 24$ 
and $32 \times 32$ lattices show that the characteristic
temperature scales reduce very slowly with increasing size.
(ii)~To consider the possibility of freezing without long range
order we compute a MC based `relaxation time' \cite{glass_temp}
$\tau_{avg}$
where $t_{max} \sim 10^4$ steps and $t'$ is the MC `time'.
If on lowering the temperature, the system
undergoes a magnetic ordering transition, then
there is a rapid growth in $\tau_{avg}$
accompanied by a growth
in the structure-factor at the
wavevector $\vec{Q}$.
The case where one observes a rapid growth
in $\tau_{avg}$
but not in the structure-factor at 
any $\vec{Q}$, suggests a glass transition
\cite{glass_temp}.
(iii)~The distribution of the magnitude of the auxiliary
moments is given by $P(m)$. Since the presence of a gap in
the electronic density of states depends on the typical
size of the ${\vec m}_i$, $P(m)$ is an important input in
inferring transport.
The mean moment $m_{avg} = \int m P(m) dm$.

The overall electronic density of states (DOS) can be 
calculated from the single particle eigenvalues, $\epsilon_n$,
in the equilibrium configurations, as:
$$
D(\omega) = \frac{1}{N} \sum_n \langle \delta(\omega - \epsilon_n) \rangle
$$

\section{RESULTS}

\subsection{Ground state}

\begin{figure}[t]
\begin{center}
\includegraphics[width=7.5cm,height=10cm]{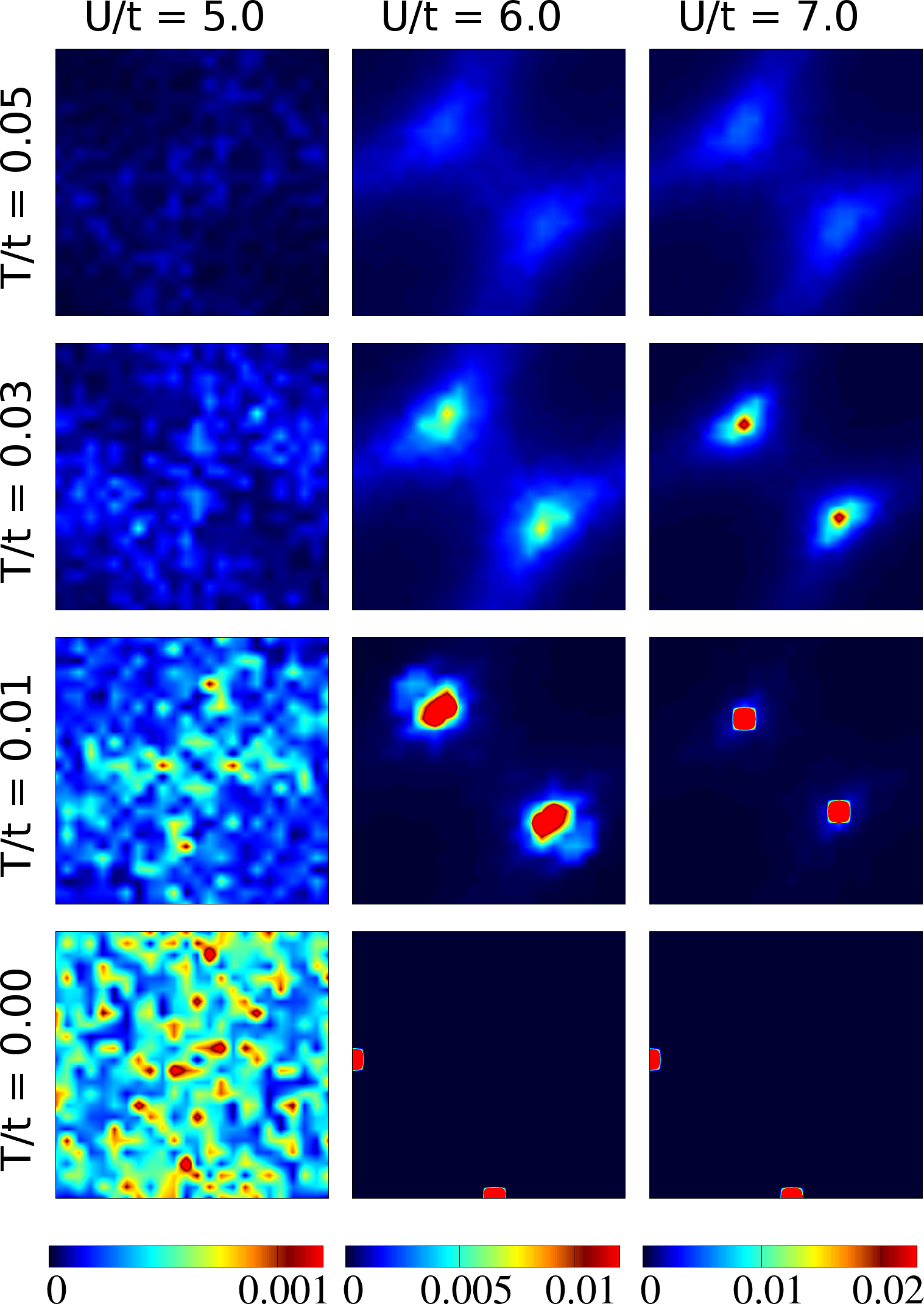}
\caption{\label{sf}
Color online: Magnetic structure factor $S(\vec{q})$
over  $q_{x}$, $q_{y}$ $\in$
$[0,2\pi]$ for different temperature and interaction strengths.
The value of $S(\vec{q})$ follows the color code.
The left column, at $U = 5t$ has no prominent peaks
at any temperature, indicating a short range correlated state.
The middle and right columns have distinct peaks
at $T=0$, at locations corresponding to flux-like state, but the
peak location shifts to that of a $120$ degree correlated state
at a small finite temperature.
}
\end{center}
\end{figure}

Let us focus on the low temperature result, at $T \sim 10^{-4}t$,
obtained via a MC on lowering $T$.
We label this as
$T=0$ in the lowest row in Fig.\ref{sf}, which shows the structure factor
versus $q_x$ and $q_y$.
The leftmost panel, typical of the window $3t < U < 5.3t$,
show no prominent features in the structure factor.
It is suggestive 
of local moments with weak and short range correlations.
We will need to look at the local moment magnitudes to fully
characterise this window.
For $U \ge 5.3t$ the structure factor shows distinct peaks at 
$\vec{Q_{F_{1}}} =$ ($\pi$,0) and 
$\vec{Q_{F_{2}}} =$ (0,$\pi$) 
with the weight at peak position increasing initially
with $U/t$ and saturating for $U/t \geq 10$. 
We call this the `flux' phase. The real space arrangement of the spins
in the `flux' phase is shown in the lower panel 
in Fig.\ref{phase_diagram}.

In our MC runs, we obtain 
the `flux' state at the lowest temperature 
only in a small window of interaction strength,
$U=[5.3t, 5.7t]$, near the IMT.
The system encounters a
`120 degree' spin arrangement 
when cooled from high temperature, and does not manage
to transit to the flux state, despite the flux state
having the lowest ground state energy
for all $U/t \geq 5.3$.
On heating up from the flux state the order
survives to a low finite $T$ and then 
the system enters the 120 degree phase.
We will discuss the results of the variational approach
and its consistency with the MC results further on.

\subsection{Finite temperature correlations}

\begin{figure}[b]
\centering{
\hspace{-.7cm}
\includegraphics[width=8.0cm,height=6.0cm]{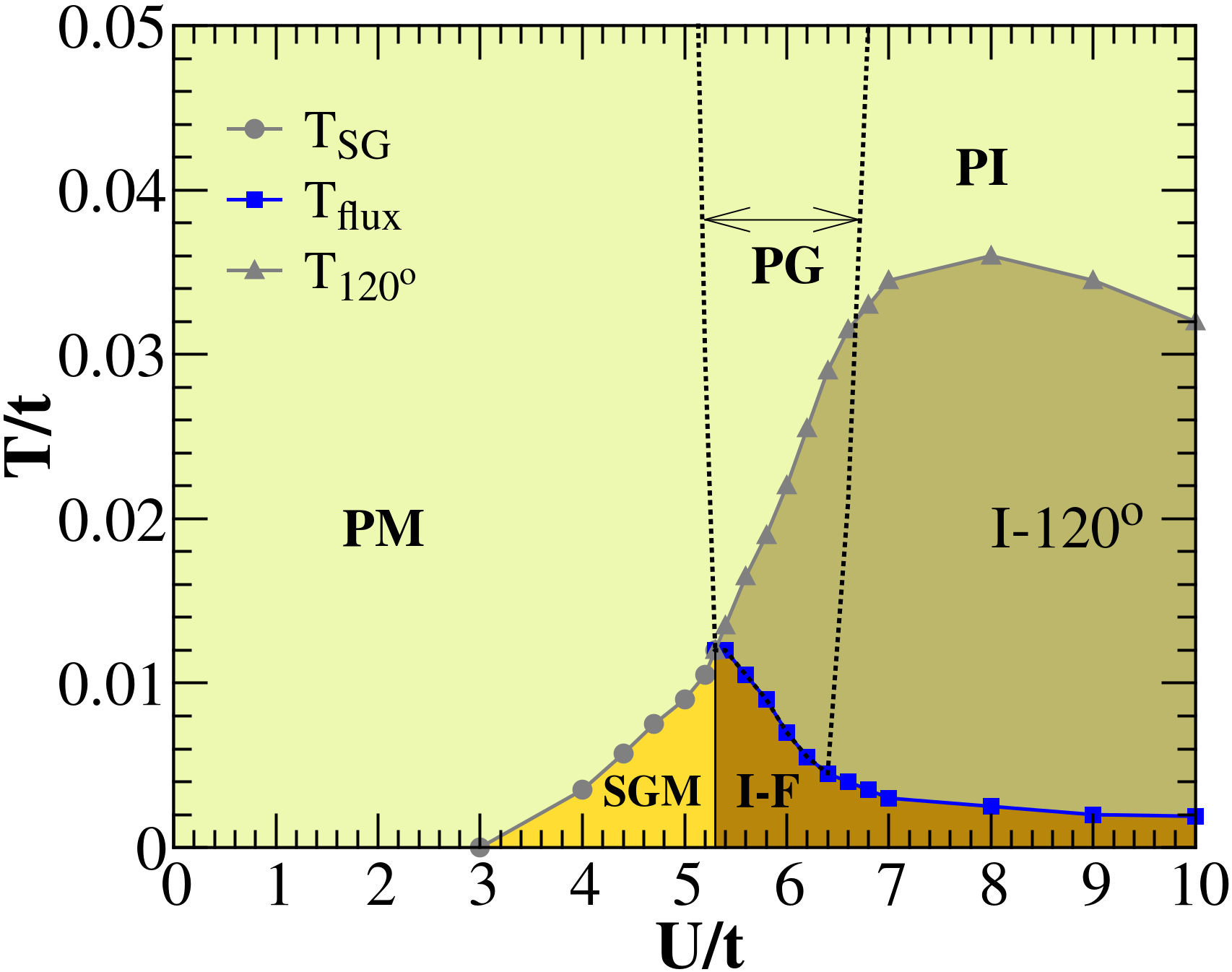}
}
\centerline{
\includegraphics[angle=0,width=3.5cm,height=3.2cm]{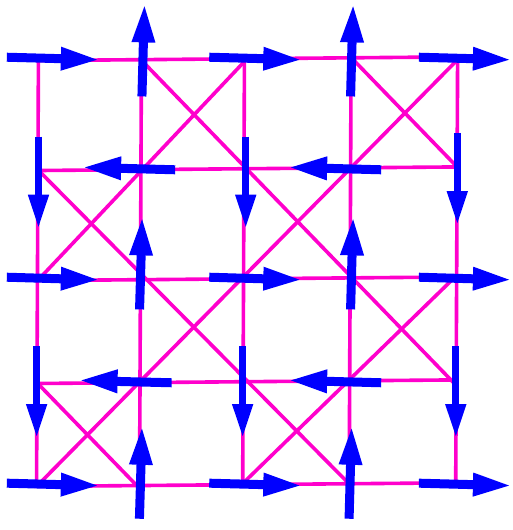}
\includegraphics[angle=0,width=3.5cm,height=3.2cm]{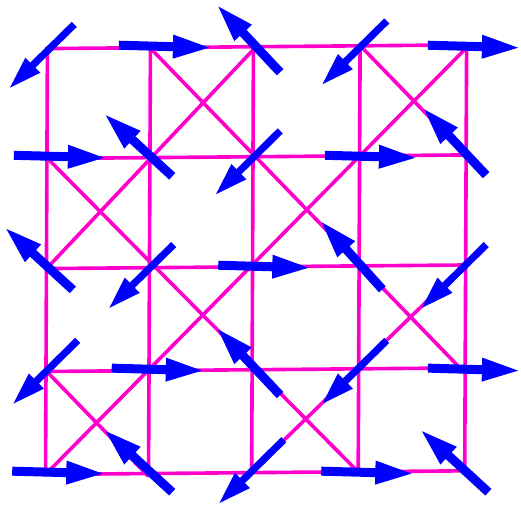}
}
\caption{\label{phase_diagram}
Color online: Top:  
The phase diagram of the checkerboard  Hubbard model
at half-filling. The ground state is a paramagnetic metal (PM)
for $U \le  3t$, a spin glass metal (SGM) for $3t < U < 5.3t$,
and an insulator with `flux' like correlations (I-F) for $U \geq 5.3t$.
With increase in temperature, the `flux' correlated phase transforms
to a $120$ degree correlated state (I-$120^{o}$) and then to 
the paramagnet (PI). The $120$ degree phase is gapped, 
except for $ 5.3t \le U \le 6.5t$, near the insulator-metal transition, 
where it shows a pseudogap (PG).
The different temperature scales shown in this panel correspond to 
magnetic correlation length growing beyond system size. 
These temperature scales are crossover scales from one
magnetic phase to another. $T_{SG}$ represents a crossover from a PM
to a SGM phase, $T_{flux}$ represents a crossover from flux phase
to $120$ degree phase and $T_{120^{o}}$ refers to a crossover
from $120$ degree state to a paramagnet.
Bottom: A schematic of the
`flux' phase (left) and the 120 degree phase 
(right) on the checkerboard lattice.}
\end{figure}

There are four broad coupling regimes in terms of 
temperature effects.
The data in Fig.\ref{sf} is for the two central regimes, 
but we discuss all the four regimes below.

(i)~At very weak coupling, $U < 3t$, there are no local
moments in the ground state. Increasing temperature {\it generates}
small moments but there are no significant spatial correlations 
between them. Since this state is fairly obvious we have not
included the result for this regime in Fig.\ref{sf}.

(ii)~At somewhat larger couplings, $3t < U < 5.3t$,
there are disordered moments in the ground state. 
These moments seem to be frozen 
on the basis of $\tau_{avg}$ estimates and the frozen state 
survives to a `spin-glass temperature' $T_{SG}$.
The left column in Fig.\ref{sf} shows the $T$ dependence of
magnetic structure factor, $S({\vec q})$ in this coupling window. 
It is clearly seen that $S({\vec q})$ is featureless in this
regime.

(iii)~For larger interaction strength, $U \sim [5.3t,8t]$,
$S({\vec q})$ shows peaks at wavevectors 
$\vec{Q_{F_{1}}} = (\pi,0)$ and $\vec{Q_{F_{2}}} = (0,\pi)$ 
in the ground state. We call this the `flux' phase. 
The amplitude at these wavevectors decrease 
with increasing $T$ and beyond a scale $T = T_{flux}$ 
new peaks appear at $\vec{Q_{T_{1}}} = (\pi/3,2\pi/3)$ 
and $\vec{Q_{T_{2}}} = (2\pi/3,\pi/3)$.
These peaks correspond to a $120$ degree 
arrangement of the moments.
This is visible in the middle and right columns in
Fig.\ref{sf}. 
The weights at the ${\vec Q}_T$ increase quickly 
with temperature, reach a maximum, and then
decrease - vanishing at  $T = T_{120^o}$. 
For $T > T_{120^o}$ there is no peak 
in $S(\vec{q})$ for any $\vec{q}$,
indicating the paramagnetic regime.
In this $U/t$ window 
$T_{120^o}$ {\it increases} with increasing $U$.

\begin{figure}[t]
\centerline{
\includegraphics[angle=0,width=9.5cm,height=4.9cm]{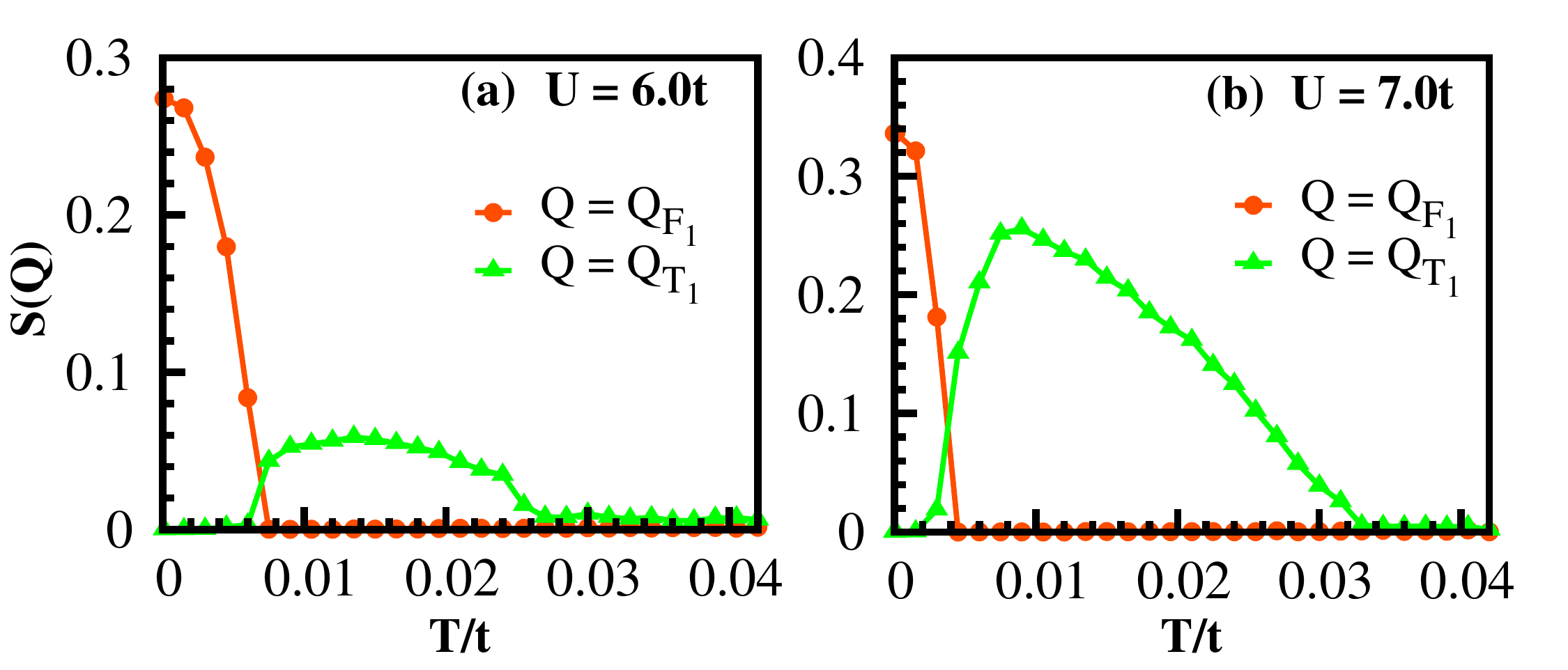}
}
\caption{\label{sf_peak}
Color online: Variation of structure factor peak amplitude
at the flux wavevector $\vec{Q_{F_{1}}} = (\pi,0)$ and
the triangular arrangement wavevector
$\vec{Q_{T_{1}}} = (\pi/3,2\pi/3)$ with temperature.}
\end{figure}

(iv)~In the asymptotically large coupling regime, $U \gtrsim 8t$,
the same sequence of flux and $120$ degree correlations 
are obtained with increasing temperature, but the characteristic
$T$  scales fall with increasing $U/t$.

The top panel in Fig.\ref{phase_diagram} shows 
the phase diagram based on the $S({\vec q})$.
Fig.\ref{sf_peak} shows the $T$ dependence of 
the structure factor peak, highlighting
the multiple thermal crossovers observed 
within our MC. 
Here $S(\vec{Q_{F_{1}}})$ corresponds to the
structure factor for flux like order, 
whereas $S(\vec{Q_{T_{1}}})$ corresponds to the 
structure factor for `$120$ degree' like order.

\subsection{Local moment distribution}

\begin{figure}[t]
\centerline{
\includegraphics[angle=0,width=11.5cm,height=8.5cm]{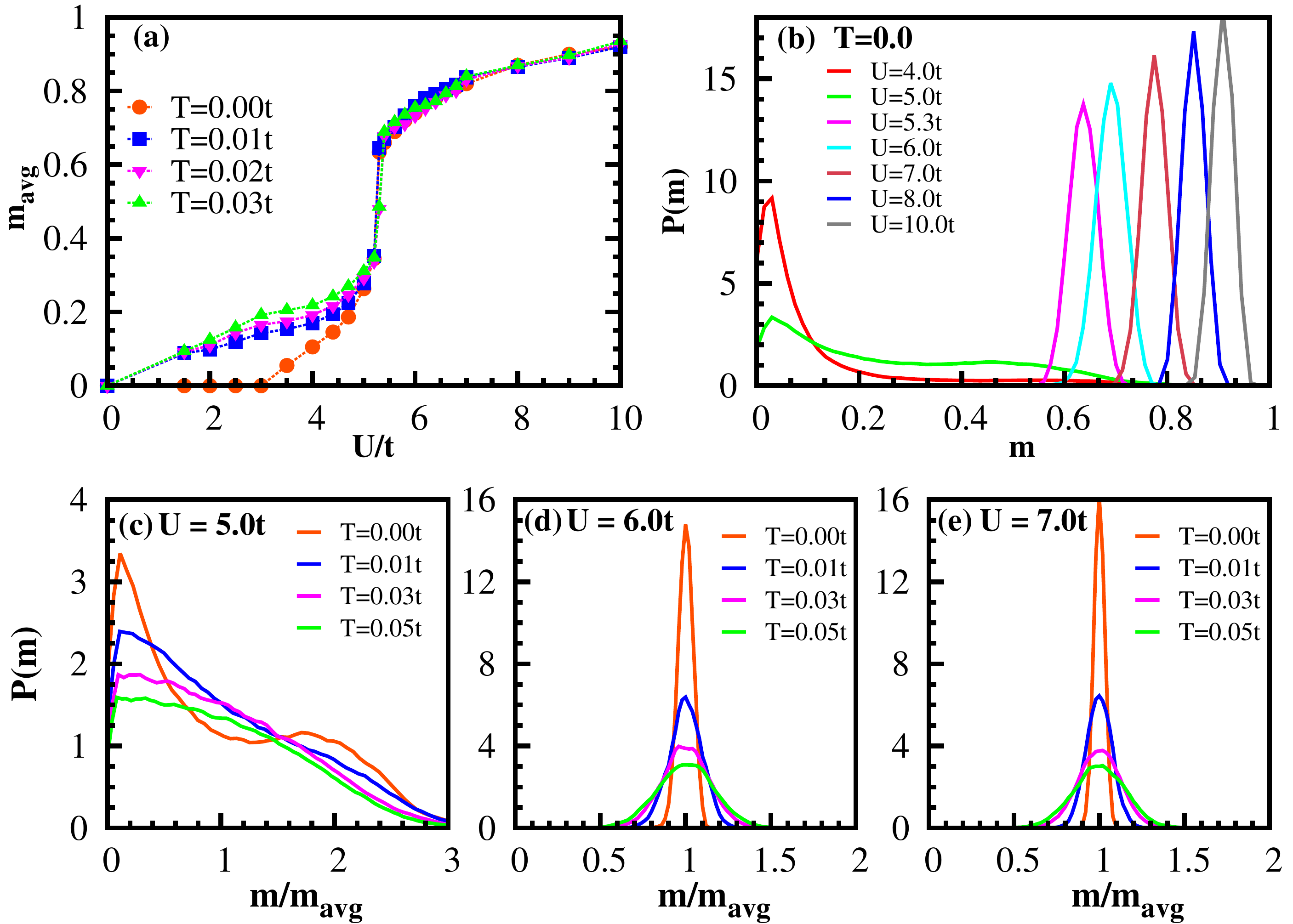}
}
\caption{\label{avg_m_pm}
Color online:
(a)~Temperature dependence of average local moment $m_{avg}$ with $U/t$.
(b)~$P(m)$ in the ground state for varying $U/t$.
(c)-(e)~Temperature dependence of $P(m)$ for $U/t = 5.0,~ 6.0,~7.0$.
}
\end{figure}

Fig.\ref{avg_m_pm}(a) shows the behavior of 
the mean local moment magnitude
$m_{avg}$ with interaction strength.
In the ground state, there is no local-moment 
for $U \le U_{c1} \sim 3t$.
There is a small average moment 
for $U_{c1} \leq U < U_{c2} \sim 5.3t$.
The average moment increases with 
interaction strength in the $[U_{c1},U_{c2}]$ window
but for $U = U_{c2}$ there is a jump 
in the average moment value. 
The average moment again increases slowly 
for $U > U_{c2}$ and saturates to $ m_{avg} = 1.0$ 
as $U/t \rightarrow \infty$.

With increase in temperature there are both
orientational and magnitude fluctuations in the
moments.
Though the average moment remains unchanged 
in the strong interaction side, it shows 
significant changes in the weak interaction side
due to the small amplitude stiffness.

Fig.\ref{avg_m_pm}(b) shows 
the $P(m)$ for the ground state. 
This evolves from a broad distribution 
in the spin glass window to essentially a 
delta function in the Mott phase.
To get a feel for the thermal fluctuations 
at different interaction strengths we have plotted 
the $P(m)$ vs $m/m_{avg}$ for $U = 5t,~6t,~7t$
and different temperatures (Fig.\ref{avg_m_pm}(c)-(e)). 
At $U=5t$ the distribution is already broad at $T=0$ 
due to the amplitude inhomogeneity in the glassy state.
The low $T$ for which the data is shown does not lead to
significant change.
At $U=6t,~7t$ the $T=0$ result is essentially a delta function 
and it broadens slightly on raising temperature. 
On the strong coupling, Mott, side the dominant fluctuations
are in the orientation of the moments, not their magnitude.
 
\subsection{Insulator-metal transition}

\begin{figure*}[t]
\begin{center}
\includegraphics[width=13cm,height=4.5cm]{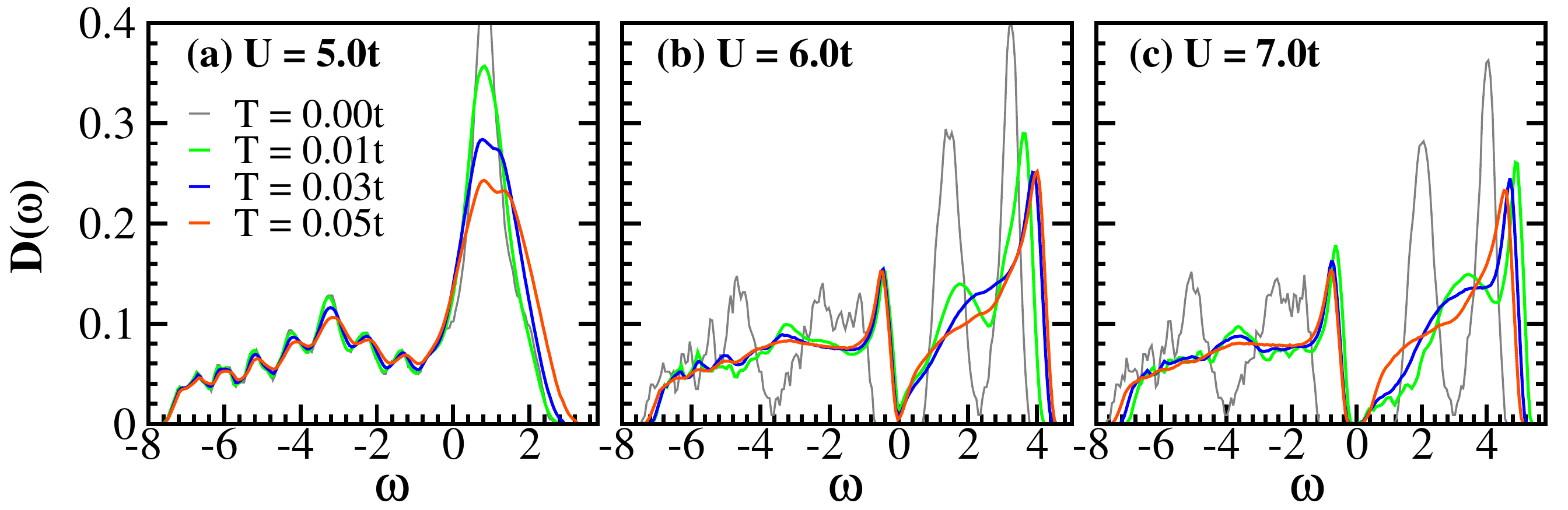}
\includegraphics[width=13cm,height=4.5cm]{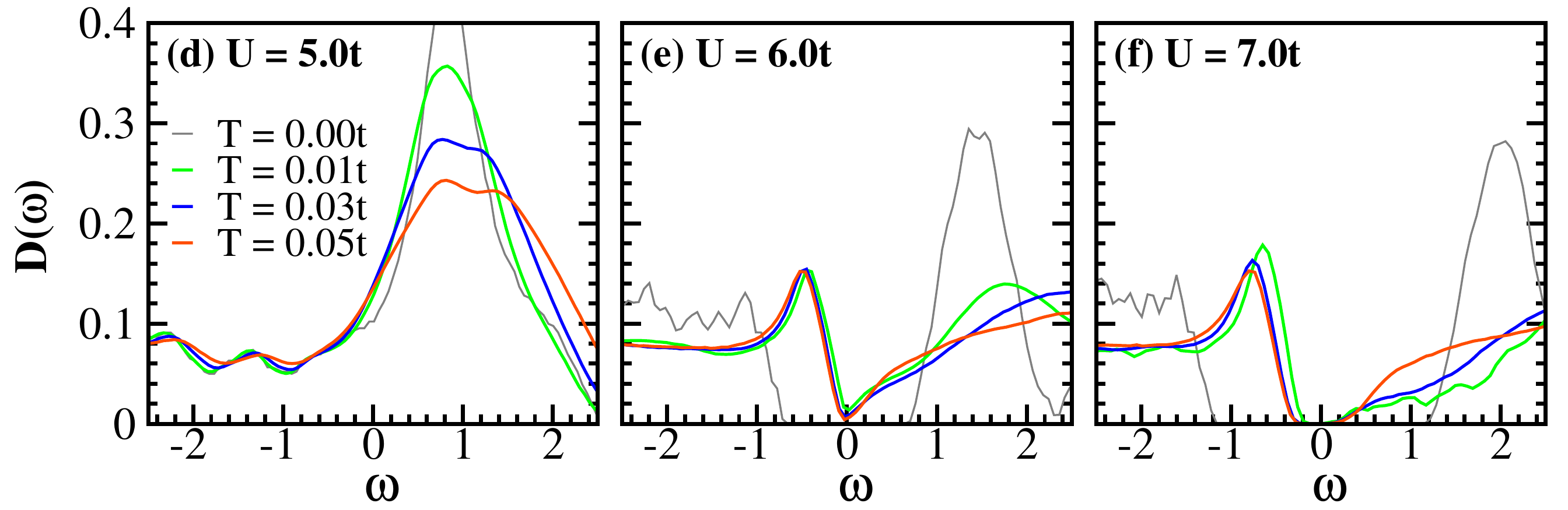}
\caption{\label{dos}
Color online:
Electronic density of states (DOS)
for $U/t = 5.0,~ 6.0,~7.0$ and varying temperature. The upper row shows
the DOS over a wide frequency window to emphasize its global features
while the lower row shows the DOS over a much smaller frequency window
centered on the Fermi level.
}
\end{center}
\end{figure*}

The first guess about the metallic or 
insulating behavior of the electrons can be
made from the single particle density of states  
$D(\omega)$.
Fig.\ref{dos} shows the density of states  
for different $U/t$  and temperatures.
For $U/t = 5.0$ the ground state has small $m_{avg}$, $ \approx 0.1$, 
and spatially disordered local moments.
These local-moments are not large enough 
to open  a gap at the Fermi energy. 
They broaden the flat tight-binding band, generating 
finite DOS 
at the Fermi energy. The system would be metallic 
in this regime.
For $U/t = 6.0$ the local-moments are 
sizable, $m_{avg} \approx 0.7$, large enough 
to open a gap in the DOS. 
Thus the system is insulating in this regime. 
With further increase in $U/t$, 
$m_{avg}$ increase monotonically saturating to $m_{avg} \approx 1.0$.

With increase in temperature the local moments 
fluctuate, both in amplitude and orientation. 
At weak coupling,  $U < U_{c2}$, 
the small fluctuating moments 
broaden the DOS feature around $\omega = 0$ 
maintaining the metallic nature. 
In the strong coupling side, $U/t > 7.0$, 
the sizable moments
maintain a gap in the DOS despite strong 
angular fluctuations. 
At intermediate coupling,  $5.0 < U/t < 6.4$, the DOS shows 
a dip at $\omega  = 0$ without any clear gap. 
We call this  `pseudogap' (PG) phase. 
The PG feature survives upto very large temperature.

The metallic or insulating 
character should actually be inferred
from a conductivity calculation.
At large $U/t$ the presence of a gap is enough
to ensure that the system would be insulating, without
having to compute the conductivity. On the small $U/t$ side
however, the situation is more complicated since we have
a disordered situation in 2D. Since the disorder is weak
and of a magnetic nature (rather than a scalar potential)
we guess that spin flip scattering would sustain a metallic
state.

\section{DISCUSSION}

\begin{figure}[t]
\centering{
\includegraphics[width=10cm,height=7.5cm,angle=0]{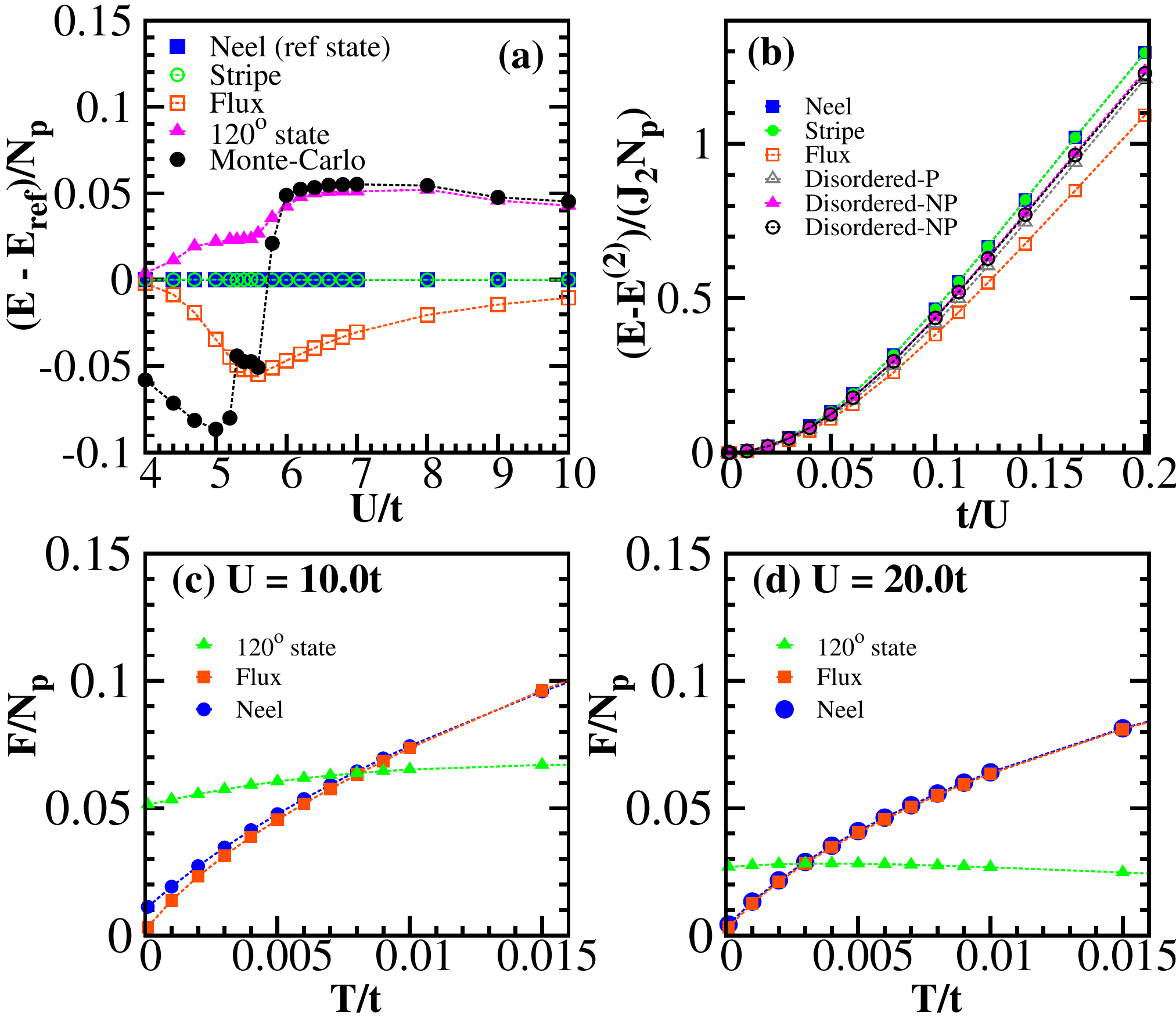}
}
\caption{\label{mc_vc_comp}
Color online: (a)~Ground state energies of
variational Neel, stripe, flux, the $120$ degree state, 
and the state obtained from Monte-carlo cooling.
Energy of the Neel state is taken as the reference energy.
(b)~Energy due to the quartic hopping 
processes (leading $t^4/U^3$ corrections).
The energy change of the Neel, $120$ degree state,
and disordered zero-plaquette-spin configurations
(planar and non-planar) are higher than the flux phase 
(they all have the same energy in the Heisenberg limit).
(c)-(d)~Free energy due to low-lying excitations on 
the variational Neel, flux, $120$ degree triangle states
at two values of $U/t$. The methodology is explained in
the text.
Notice that the free energy of the $120$ degree
state falls below that of the flux phase at a temperature that
reduces with increasing $U/t$.
$N_p$ is the number of plaquettes on the lattice.
}
\end{figure}

\subsection{Origin of the magnetic phases}

Let us describe the variational scheme that we have used to
complement the Monte Carlo and then move on to the analysis
of the magnetic phases observed in Fig.\ref{sf}.
We set up trial states using spiral spin configurations,
${\vec  m}_i =
m_0(cos(\vec{Q}.\vec{R_i}),sin(\vec{Q}.\vec{R_i}),0)$,
with uniform magnitude $m_0$ and wave-vector $\vec{Q}$ as
variational parameters, and minimised the energy of $H_{eff}$ at
half-filling. This differs slightly from the real situation
where the ${\vec  m}_i$'s have some amplitude inhomogeneity.
We also included several `non spiral' configurations,
notably the flux, that satisfy the 
local constraint of vanishing plaquette spin.

Comparing the minimum energy obtained via variational calculation
with that from the MC cooling run, see Fig.\ref{mc_vc_comp}(a), 
confirms that the {\it flux} state is the ground state for $U \geq U_{c2}$.
However the inhomogeneous small moment `spin glass'
phase obtained by our Monte-Carlo cooling 
indeed turns out to be the lowest energy state 
for $U_{c1} < U < U_{c2}$, lower than the trial periodic
configurations.

How do we understand the magnetic phases?  It is
helpful to consider three separate regimes: 
(i)~The $U/t \gg 1$ window 
where only the leading exchange term $J_2 \sim t^2/U$ 
is relevant, and the moment size $m_{avg} \sim 1$.
(ii)~Intermediate $U/t$, down to $U \sim U_{c2}$,
where the moment size is still 
large but higher order spin-spin couplings, 
in particular $J_4 \propto t^4/U^3$, 
is significant. Finally 
(iii)~the `weak coupling' end, $U \sim 3t$, 
where the moment is small and 
it is more appropriate to expand about the
band limit rather than the Mott state.
Let us consider the ground state and 
thermal effects in succession.

{\bf Ground state:}
In the strong coupling limit, we can write 
an effective magnetic Hamiltonian on this lattice 
by tracing out the fermions order by order in $t/U$. 
This gives us
\begin{eqnarray}
H_{eff}\{{\bf m}\} & = & E^{(2)}\{{\bf m}\} 
+ E^{(4)}\{{\bf m}\} + ... \cr
~~ \cr
E^{(2)}\{{\bf m}\} & =  & J_2
\sum_{\boxtimes} \sum_{\langle i,j 
\rangle} ({\bf m}_i . {\bf m}_j -1 )
~~ \cr
E^{(4)}\{{\bf m}\} & = & E^{(4)}_{ \boxtimes }\{{\bf m}\} +
 E^{(4)}_{coup}\{{\bf m}\} \cr
~~ \cr
E^{(4)}_{\boxtimes}\{{\bf m}\} & 
= & J_4 \sum_{\boxtimes} 
[~~~ 
\sum_{\langle i,j \rangle} 
\{ \frac{5}{2} ({\bf m}_i . {\bf m}_j)^2 
- {\bf m}_i . {\bf m}_j -\frac{3}{2} \}\cr
~~ \cr
& ~ &~~~~~~~~+\sum_{\langle i,j,k \rangle} 
\{ 5({\bf m}_i . {\bf m}_j)({\bf m}_j . {\bf m}_k)
- 3({\bf m}_i . {\bf m}_k)   \} \cr
~~ \cr
& ~ &~~~~~~~~+\sum_{\langle i,j,k,l \rangle} 
\{ 5({\bf m}_i . {\bf m}_j)({\bf m}_k . {\bf m}_l)
+5({\bf m}_i . {\bf m}_l)({\bf m}_j . {\bf m}_k) \cr
~~ \cr
& ~ &~~~~~~~~~~~~~~~~~~~~
+5({\bf m}_i . {\bf m}_k)({\bf m}_j . {\bf m}_l)  \}]
~~\cr
E^{(4)}_{coup}\{{\bf m}\} & \sim & 
J'_4 \sum_{i\in \boxtimes_1 ,j\in \boxtimes_2} {\bf m}_i . {\bf m}_j 
+J''_4 \sum_{i\in \boxtimes_1 ,j\in \boxtimes_2}^{k\in 
\boxtimes_1 \cap \boxtimes_2} 
({\bf m}_i.{\bf m}_k)({\bf m}_j.{\bf m}_k)... \nonumber
\end{eqnarray}
where $\boxtimes$ represents the crossed plaquette.
$E^{(2)}\{{\bf m}\}$ corresponds to the second 
order perturbation energy
with $J_2 = t^2/U $, and  $E^{(4)}\{{\bf m}\}$ 
corresponds to the fourth order perturbation 
energy. $E^{(4)}_{\boxtimes}$ describes the 4th
order terms 
within a crossed-plaquette,
while $E^{(4)}_{coup}$ describes the interplaquette terms
with a common corner shared site.
We have found ~$J_4 = \frac{t^4}{U^3}$
and $J'_4, J''_4 \sim  O(\frac{t^4}{U^3})$.

In regime (i), one would drop the $E^{(4)}\{{\bf m}\}$ term 
and obtain a classical Heisenberg model.
On the checkerboard lattice the Heisenberg interaction 
can be written as the sum of squares of the total spin
on each plaquette, ${\vec S}_P = \sum_i {\vec m}_i$, 
where the sum is over spins in individual crossed plaquettes. 
This feature arises due to the `fully connected' nature 
of the crossed plaquettes, 
which are essentially tetrahedra, and is true of the
pyrochlore lattice as well. 
The minimum energy corresponds to all ${\vec S}_P =0$
but there is a macroscopic degeneracy 
in the number of ways this can be satisfied.
In such degenerate situations 
thermal fluctuations sometime select out
collinear ordered configurations,
due to the entropy gain \cite{villain}. 
For the checkerboard lattice
it seems that  free energy barriers are small
and the thermal `order-by-disorder'
mechanism does not select out an ordered state. 
The classical Heisenberg limit, $U/t \rightarrow
\infty$, is disordered at all temperatures as in the
pyrochlore lattice \cite{chkb_cHAF}.
For the Hubbard model the actual spins are $S=1/2$ and
not classical, and a $1/S$ expansion about the classical
limit suggests that a `quantum order by disorder' mechanism
selects a valence bond solid (VBS) ground state
\cite{chkb_HAF-sw1,chkb_HAF-sw2,chkb_HAF-sw3}.

In regime (ii) the contribution 
of $E^{(4)}\{{\bf m}\}$
becomes important. This can be seen as follows.
Various configurations satisfying the local constraint,
$\sum_{i \in \boxtimes} {\bf m}_{i} = 0$
have equal $E^{(2)}\{{\bf m}\}$.
However the Hubbard energy for these different
configurations are found to be different.
Thus the crucial difference to the Hubbard energy 
is dominated by the $E^{(4)}\{{\bf m}\}$ contribution.
We believe, these multi-spin exchange interactions
are responsible in modifying the magnetic ground state
away from the Heisenberg limit.
In figure \ref{mc_vc_comp}.(b), 
we show this contribution as
$(E - E^{(2)}\{{\bf m}\})/J_2 \approx 
(t/U)^2 f({\bf m}_i,{\bf m}_j,{\bf m}_k,{\bf m}_l)$
with the expected quadratic behaviour as 
$t/U \rightarrow 0$ (or $U/t \rightarrow \infty$).
Its also seen that the flux state has the largest lowering 
of energy to ${\cal O}(t^4/U^3)$.
This trend persists down to $U \sim U_{c2}$.
The strong coupling expansion in $t/U$ ceases to be useful
once the Mott gap closes.

(iii)~At weaker coupling, $U \sim [3t,5.3t]$, the system is
gapless and the moments are small. The state as $U \rightarrow 3t$
is better understood as an instability in the tight binding
band, controlled by the susceptibility $\chi_0({\vec q})$.
In non frustrated lattices this usually has a prominent peak
at some ${\vec q} = {\vec Q}$, and the condition $1 - U \chi_0({\vec Q})
=0$, defines the onset of ordering at $U_c$. We computed $\chi_0({\vec q})$
for the checkerboard lattice and discovered that it is essentially
featureless. No specific wavevector is selected out, so when the
local moments form they encounter competing interactions in real
space. This, within our scheme, appears to lead 
to a spin frozen state.

{\bf Finite temperature:}
The finite temperature state is dictated by the free-energy 
of the possible low energy ordered 
configurations. While the
entropy difference around different ordered configurations is
not large enough to stabilise long range order in the
Heisenberg limit, we wanted to
check how the situation is modified in the Hubbard model.
We tried a rather crude free energy  
estimate to gain some insight since the 
explicit ${\vec m}_i$ based model is not available at 
intermediate coupling. 

We considered a homogeneous ordered state 
and chose a reference site ${\vec R}_0$. 
We create a single spin `fluctuation' on the ordered state 
by giving angular twists to ${\vec m}_{R_0}$ 
without disturbing the other ${\vec m}_i$s.
We calculate the energy cost for these fluctuations 
with respect to the ordered state by using the Hubbard model. 
This process was repeated for
random twists distributed uniformly on the surface of a sphere
and for different temperatures. An averaging over the reference site
also had been taken into account. 
The density of states 
of these single spin excitation energies allows us to 
roughly estimate the free-energy.
It is expected that the states with a high density of low
energy excitations would be preferred since they have
the largest entropy.

Our results Fig.\ref{mc_vc_comp}(c),(d) show that 
at intermediate temperature the checkerboard 
Hubbard model prefers a $120$ degree correlated state (which does not 
satisfy the plaquette constraint) due to entropic reasons. 
The $120$ degree state, however, has a higher internal energy than
the flux state and loses out to it at a lower $T$.

\begin{figure}[t]
\begin{center}
\includegraphics[width=7.0cm,height=5.3cm]{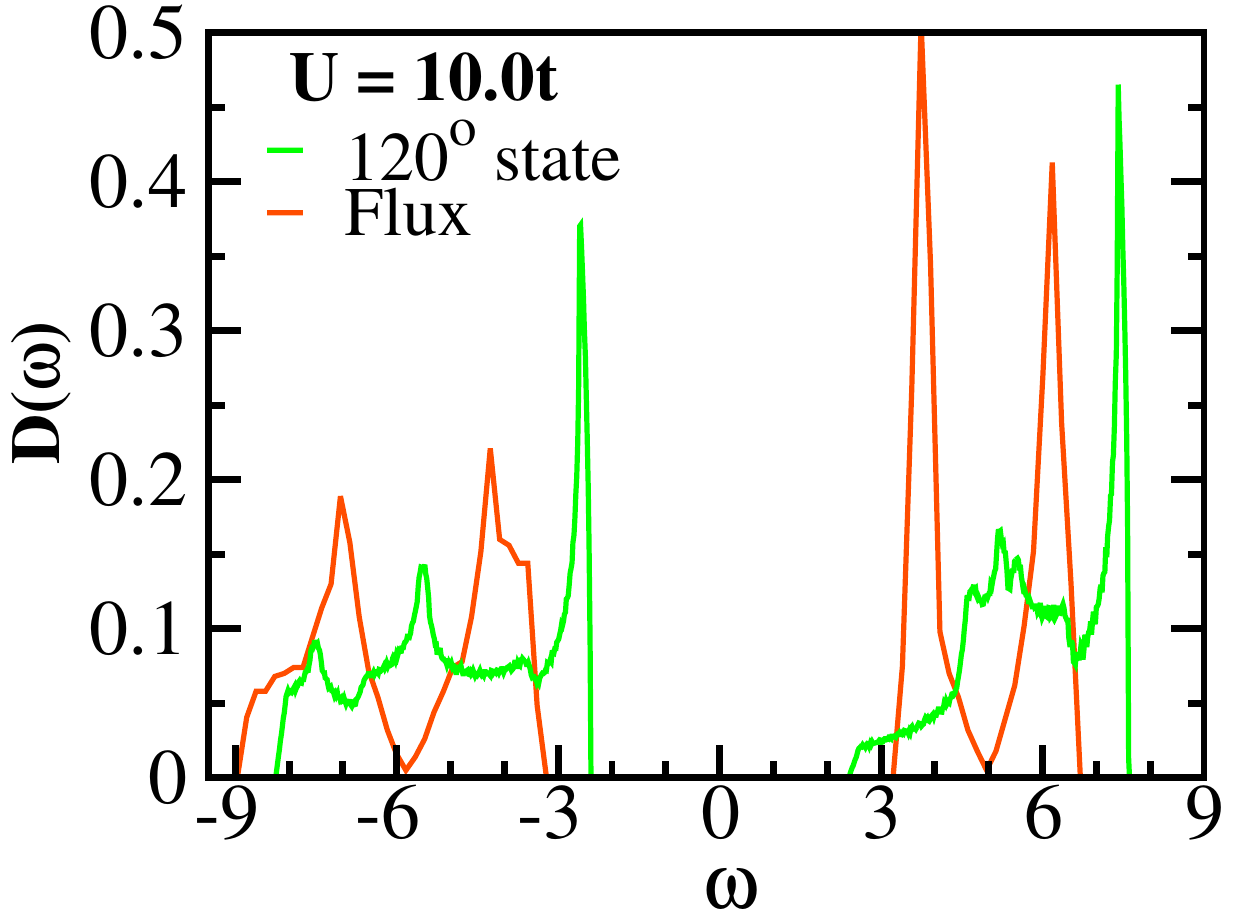}
\caption{\label{dos_ideal}
Color online:
Electronic DOS in the ideal flux and $120$ degree spin configurations.
For the same $U/t$ the $120$ degree state has a smaller gap than 
the flux phase.} 
\end{center}
\end{figure}

We would like to point out the differences
of our results in the present study,
from the earlier studies \cite{mit_yoshioka} 
of the Mott transition on the checkerboard lattice.
We address it in terms of the validity of our approximation 
in the $U-T$ plane based on $T=0$ and finite $T$ results.
(i)At $T=0$, these three features are noteworthy. 
(a)~The $U_c$ for the Mott transition: we obtain $U_c \sim 5.3t$, 
the only other value we know in the literature is $U_c \sim 6.7t$.
These are in the same ballpark. 
(b)~The large $U$ state: we obtain a flux like state 
while the stdudy \cite{mit_yoshioka} 
uses a more sophisticated approach 
to obtain a plaquette singlet state. Within our 
approach we believe the singlet state can be accessed only
if we include quantum fluctuations of the ${\bf m}_i$'s. 
Both the plaquette singlet and 
the flux state lift the classical degeneracy of
the Heisenberg limit but through different mechanisms. 
(c)~The low $U$ metallic spin glass would be susceptible to quantum
fluctuations, since it is a gapless state, and the Hartree-Fock
result is likely to be modified in a full theory.\\
(ii)Finite $T$: While our $T=0$ results have the limitation
of being Hartree-Fock, increasing $T$ brings into play the
fluctuations that were suppressed at $T=0$. In fact as $T$ 
grows these classical fluctuations dominate over the quantum
fluctuations and dictate the magnetic correlations and the
electronic properties. To our knowledge, this aspect of
checkerboard Mott physics has not been explored before.

\subsection{Magnetic impact on electronic properties}

Within our framework the electronic properties are
dictated by the behaviour of the local moments,
which in turn is decided by the electrons. We wish
to establish a more quantitative connection between the
magnetic order and the electronic DOS in some limiting cases.

For $T=0$ and  $U \leq U_{c1}$ 
there are no local moments and the system
is described by the tight-binding model. 
On the checkerboard structure this leads to a 
flat band at the upper band edge.
For $U_{c1} \leq U < U_{c2}$ small moments 
show up,  modifying the tight-binding 
DOS by broadening the flat band.
For $U \geq U_{c2}$ the moments are sizable and 
they open a gap in the DOS. The `flux' like order
has a unique 4-peak structure in the DOS with 
a wide gap around the Fermi level.
The $120$ degree `triangle' phase shows pseudogap 
in the intermediate interaction window and 
has a gapped phase at strong interaction side. 
The specific behavior of the DOS in these 
magnetic ordered states can be understood as follows.

In the `flux' phase, the local-moment at any lattice site 
${\vec R}_i$ can be parametrized as $\vec{m_{i}} = (m_{ix}, m_{iy}, 0)$ 
where
\begin{eqnarray}
m_{ix}& = &\frac{m}{2} 
(e^{i\vec{Q_{F_{1}}}.\vec{R_{i}}} + 
e^{i\vec{Q_{F_{2}}}.\vec{R_{i}}}) \cr
m_{iy} & = &\frac{m}{2} 
(e^{i\vec{Q_{F_{2}}}.\vec{R_{i}}} 
- e^{i\vec{Q_{F_{1}}}.\vec{R_{i}}}) \nonumber
\end{eqnarray}

This magnetic ordering leads to two distinct energy levels
at $\pm\frac{U}{2}$, each with a two-fold degeneracy.  
As electrons move on this magnetic background, they further
split to bands 
 $\pm (U/2) \pm t \sqrt{2 + cos 2k_{x} + cos 2k_{y}}$. 
Thus the electron motion in the `flux' phase 
gives rise to the unique 4-peak structure in the DOS.
The minimum gap in the DOS for this state
is $U-4t$. 

In the `120 degree' phase, the local-moment at any lattice site 
${\vec R}_i$ can be parametrized as $\vec{m_{i}} = (m_{ix}, m_{iy}, 0)$ 
where
$ m_{ix} =  m cos({2\vec{Q_{T_{1}}}.\vec{R_{i}}})$,~
$m_{iy} =  m sin({2\vec{Q_{T_{1}}}.\vec{R_{i}}})$.
This phase also has two-fold degenerate energy levels
at $\pm\frac{U}{2}$.  
Itinerant electrons on this magnetic background lift 
the degeneracy by $ t[ -g(k_{x},k_{y}) \pm h(k_{x},k_{y})]/4$
where
\begin{eqnarray}
g(k_{x},k_{y}) & = &
[4 cos (k_{x}-k_{y}) + cos (k_{x}+k_{y}) + \sqrt{3} sin (k_{x}+k_{y})] \cr
 h^{2}(k_{x},k_{y}) &= & 
[26 + 8 cos (2k_{x}+2k_{y}) - cos (2k_{x}-2k_{y}) -8 cos (2k_{x}) \cr 
&&  - 8 cos (2k_{y}) - 8 cos (k_{x}+k_{y}) + 16 cos (k_{x}-k_{y}) \cr
&&  +\sqrt{3} sin (2k_{x}-2k_{y}) -8\sqrt{3} sin (k_{x}+k_{y}) -8\sqrt{3} sin (2k_{x})] 
\nonumber
\end{eqnarray}

Thus the motion of electrons in the `120 degree' phase retains the
upper and lower Hubbard band features without undergoing 
any further splitting of bands (unlike the `flux' phase).
The minimum gap in the DOS for this state
is $U-t[(h+g)_{max} + (h-g)_{max}]/4$. We observe that 
in the Brillouin zone $[(h+g)_{max} + (h-g)_{max}] > 16$.
Thus the gap for the ideal `flux' phase is always larger
than the ideal `120 degree' phase for same interaction strength
Fig.\ref{dos_ideal}.

\section{CONCLUSION}

We have studied the single band Hubbard model 
at half-filling on the checkerboard lattice.
The Hartree-Fock ground state is non magnetic upto an
interaction strength $U_{c1}$, then  a small moment 
spin glass upto $U_{c2}$, and a `flux' ordered
state beyond. The Mott transition, associated
with a gap opening in the density of states, occurs
at $U_{c2}$.
The presence of order differentiates this lattice of
corner shared `tetrahedra' from its
three dimensional counterpart, the pyrochlore lattice,
which remains disordered at all interaction strengths.
A static auxiliary field based Monte Carlo
provides an estimate of the temperature window over which
the magnetic correlations survive. Strikingly, we observe
that the flux order is replaced by a `120 degree'
correlated spin arrangement at intermediate temperature
before all order is lost. 
We provide an entropic argument for this effect.

{\it Acknowledgements:} 
We acknowledge use of the HPC  Clusters at HRI.
NS appreciates discussing a variational calculation
result with Ajanta Maity, and 
learning about a vector graphics language 
({\it Asymptote}) from Rajarshi Tiwari.
PM acknowledges support from an  
Outstanding Research Investigator Award of
the Department of Atomic Energy-Science Research Council (DAE-SRC)
of Government of India.

\end{document}